\documentclass[a4paper,11pt]{article}
\usepackage{jcappub} 
\usepackage{lineno}

\usepackage[utf8]{inputenc}
\usepackage{epsfig,amsfonts,amsthm}
\usepackage{epstopdf}
\usepackage{amsfonts,amsmath,amssymb}
\usepackage{gensymb}
\usepackage{latexsym}
\usepackage{bbm}
\usepackage{array}
\usepackage{comment}
\usepackage{hyperref}
\usepackage{color}
\usepackage{pb-diagram}
\usepackage{slashed}
\usepackage{multirow}
\usepackage{cancel}
\usepackage{placeins}                       
\usepackage{diagbox}
\usepackage{soul}

\title{COSINUS model-independent sensitivity to the DAMA/LIBRA dark matter signal}

\author[a]{G.~Angloher}
\affiliation[a]{Max-Planck-Institut f\"ur Physik, 85748 Garching - Germany}

\author[a]{M.~R.~Bharadwaj}

\author[b,c]{A.~B\"ohmer}
\affiliation[b]{Institut f\"ur Hochenergiephysik der \"Osterreichischen Akademie der Wissenschaften, 1050 Wien - Austria}
\affiliation[c]{Atominstitut, Technische Universit\"at Wien, 1020 Wien - Austria}

\author[b,c]{M.~Cababie}

\author[d,e]{I.~Colantoni}
\affiliation[d]{Consiglio Nazionale delle Ricerche, Istituto di Nanotecnologia, 00185 Roma, Italy}
\affiliation[e]{INFN - Sezione di Roma, 00185 Roma - Italy}

\author[f,e]{I.~Dafinei}
\affiliation[f]{Gran Sasso Science Institute, 67100 L'Aquila - Italy}

\author[f,g]{N.~Di~Marco}
\affiliation[g]{INFN - Laboratori Nazionali del Gran Sasso, 67010 Assergi - Italy}

\author[a]{ C.~Dittmar }

\author[c,h]{L.~Einfalt}
\affiliation[h]{Consiglio Nazionale delle Ricerche, Istituto di Nanotecnologia, 00185 Roma, Italy}

\author[g]{F.~Ferella}

\author[f,e]{F.~Ferroni}

\author[b]{S.~Fichtinger}

\author[g,i]{A.~Filipponi}
\affiliation[i]{Dipartimento di Scienze Fisiche e Chimiche, Universit\`a degli Studi dell'Aquila, 67100 L'Aquila - Italy}

\author[a]{T.~Frank}

\author[b]{M.~Friedl}

\author[j]{Z.~Ge}
\affiliation[j]{SICCAS - Shanghai Institute of Ceramics, 201899 Shanghai - P.R.C}

\author[k]{M.~Heikinheimo}
\emailAdd{matti.heikinheimo@helsinki.fi}
\affiliation[k]{Helsinki Institute of Physics, 00014 University of Helsinki - Finland}

\author[a]{M.~N.~Hughes}

\author[k]{K.~Huitu}

\author[a]{M.~Kellermann}

\author[b,c]{R.~Maji}

\author[a]{M.~Mancuso}

\author[f,g]{L.~Pagnanini}

\author[a]{F.~Petricca}

\author[g]{S.~Pirro}

\author[a]{F.~Pr\"obst}

\author[g,i]{G.~Profeta}

\author[g]{A.~Puiu}

\author[b,c]{F.~Reindl}

\author[a]{K.~Sch\"affner}

\author[b,c]{J.~Schieck}

\author[b,c]{P.~Schreiner}

\author[b,c]{C.~Schwertner}

\author[a]{K.~Shera}

\author[a]{M.~Stahlberg}

\author[k]{A.~Stendahl}
\emailAdd{alex.stendahl@helsinki.fi}

\author[g,l]{M.~Stukel}
\affiliation[l]{SNOLAB, P3Y 1N2 Lively - Canada}

\author[g,m]{C.~Tresca}
\affiliation[m]{CNR-SPIN c/o Dipartimento di Scienze Fisiche e Chimiche, Universit\`a degli Studi dell'Aquila, 67100 L'Aquila - Italy}

\author[b,n,o]{F.~Wagner}
\affiliation[n]{Department of Physics, ETH Zurich, CH-8093 Zurich, Switzerland}
\affiliation[o]{ETH Zurich - PSI Quantum Computing Hub, Paul Scherrer Institute, CH-5232 Villigen, Switzerland}

\author[j]{S.~Yue}

\author[a,b]{V.~Zema}

\author[j]{Y.~Zhu}

\collaboration{The COSINUS Collaboration}

\abstract
{COSINUS is a dark matter direct detection experiment using NaI crystals as cryogenic scintillating calorimeters. If no signal is observed, this will constrain the dark matter scattering rate in sodium iodide. We investigate how this constraint can be used to infer that the annual modulation signal observed in the DAMA/LIBRA experiment cannot originate from dark matter nuclear recoil events, independently of the dark matter model. We achieve this by unfolding the DAMA modulation spectrum to obtain the implied unquenched nuclear recoil spectrum, which we then compare to the expected COSINUS sensitivity. We find that assuming zero background in the signal region, a 1$\sigma$, 2$\sigma$ or 3$\sigma$ confidence limit exclusion can be obtained with 57, 130 or 250 kg day of exposure, respectively. A simple background model indicates that in the presence of background, the exposure requirements may increase by $\sim30\%$.}

\begin{document}
\maketitle

\section{Introduction}
\label{sec:intro}
The search for the dark matter (DM) particle is among the most important quests in modern particle physics and astrophysics. Several experiments based on absolute event counts have observed no significant excess above the expected background levels, resulting in strong constraints on the DM scattering cross section with atomic nuclei \cite{SuperCDMS:2017nns,CRESST:2019jnq,PandaX-II:2020oim,XENON:2023cxc,LZCollaboration:2024lux} and electrons \cite{PandaX:2022xqx,XENON:2022ltv,LZ:2023poo}. On the contrary, the search strategy adopted in the DAMA/LIBRA experiment is based on the expected annual modulation feature in the DM event rate instead of the absolute event counts. DAMA/LIBRA does observe such modulation in the recorded event rate with overwhelming statistical significance \cite{Bernabei:2019ajy,Bernabei:2021kdo,Bernabei:2022ath}. Two other annual modulation experiments using Sodium Iodide ($\rm{NaI}$) scintillators, ANAIS and COSINE-100, however do not observe modulation of the event rate \cite{COSINE-100:2025eyc}, ANAIS alone being incompatible with the DAMA annual modulation signal at a $4\sigma$ confidence level \cite{Amare:2025dfq}. For a review of the direct detection experiments and results see e.g. \cite{Billard:2021uyg,Akerib:2022ort}.

While the DM-electron scattering cross section is constrained almost independently of the target atoms via electron recoil searches, the comparison between different elements in nuclear recoil searches is more involved due to the composite nature of the nuclear target. Therefore, constraining the nuclear recoil interpretation of the purported DAMA/LIBRA DM signal is our main interest. DAMA/LIBRA measures the recoil energy via the scintillation light produced in a recoil event. To convert the observed light to recoil energy, a calibration with sources of known energy is required. This is typically achieved with electromagnetic probes that induce electron recoil events. However, the light yields per unit recoil energy in nuclear and electron recoil events are different, related by the quenching factor ($Q\!F$). This factor might have nontrivial dependence on e.g., the recoil energy or the ${\rm Tl}$-dopant concentration of the scintillator crystal, see e.g. \cite{Cintas:2024pdu}. Therefore, if the scintillation events are interpreted as nuclear recoils, the calibration of the nuclear recoil energy spectrum is subject to uncertainty related to the determination of the quenching factor, in addition to any other measurement uncertainties.

COSINUS, on the other hand, will have a more direct access to the recoil energy via the calorimetric measurement in the phonon channel. Therefore we will work under the assumption that the COSINUS data, in the absence of excess events, will provide a limit for the DM-nucleus scattering rate in $\rm{NaI}$ as a function of the true nuclear recoil energy. The purpose of this work is to investigate how this constraint can be effectively used to exclude the DM nuclear recoil interpretation of the DAMA/LIBRA annual modulation signal. Such analysis has been performed in \cite{Kahlhoefer:2018knc} utilizing a single DAMA/LIBRA modulation energy bin. Here we extend this analysis, making use of the reported energy spectrum of the modulation, independent of the particle physics or astrophysics of DM, while considering the uncertainties related to the energy calibration of the DAMA/LIBRA measurement. 

\section{The COSINUS Experiment}

The Cryogenic Observatory for SIgnatures seen in Next-generation Underground Searches (COSINUS) \cite{Angloher:2016ooq} has been developing a cryogenic scintillating calorimeter with a NaI target crystal. This detector simultaneously reads a phonon/heat and a scintillation light signal for every particle interaction in the crystal which provides crucial advantages compared to the typical scintillation-light-only readout of NaI crystals: Firstly, a precise and particle-type-independent measurement of the true deposited energy in the NaI crystal and, secondly, event-by-event particle discrimination for electron recoils (dominant background) and nuclear recoils (sought-for signal). These features were recently demonstrated in a prototype measurement \cite{COSINUS:2023qwi,COSINUS:2023kqd}.

In parallel with the detector development, COSINUS is constructing a low-background underground facility at the Laboratori Nazionali del Gran Sasso (LNGS). The core of the facility is a so-called dry cryostat with a base temperature below 10\,mK placed in the center of a cylindrical water tank (7\,m height, 7\,m diameter) for passive shielding and active vetoing of radiogenic and cosmogenic background radiation \cite{COSINUS:2021bdj,Angloher:2024pbw}. COSINUS will start its first run within 2025 with eight detector modules with a target mass of 30\,g per module and a nuclear recoil threshold goal of 1\,keV, which corresponds to a baseline resolution of $\sigma=0.2$\,keV. The target exposure for the first run is 100\,kg day. For the second run, COSINUS foresees making use of the full number of installed readout channels and operating 24 detector modules to collect an exposure of $\mathcal{O}$(1000\,kg day) within a measurement time of 2-3 years. More details on the experimental program and the projected DM sensitivities may be found in \cite{Angloher:2025shf}. The COSINUS facility is brand new with no measured background level so far. Even though simulation-based estimates of the background levels exist \cite{COSINUS:2021bdj,Fuss:2022gbo}, we focus the discussion in this paper to the background-free case, and postpone a detailed analysis in the presence of background to a future work. We comment on the effect of the expected background at the end of section \ref{Analysis} and on a background dominated worst case scenario at the end of section \ref{sec:resolution}.

\section{Analysis procedure}
\label{Analysis}
\begin{figure}
	\begin{center}
		\includegraphics[width=0.5\linewidth]{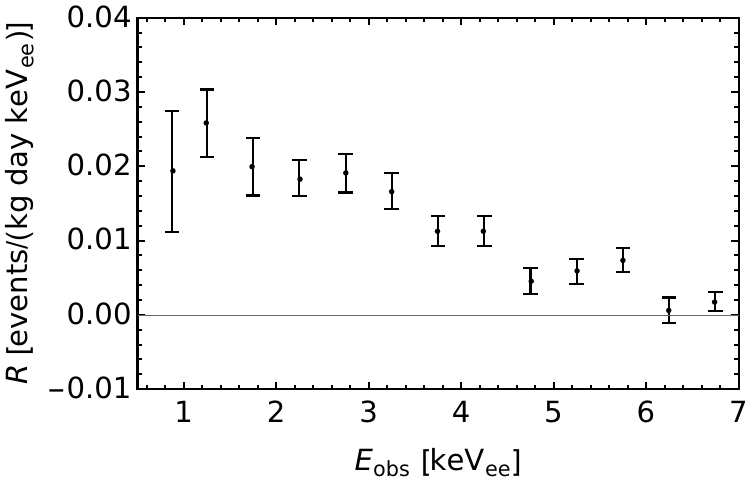}
		\caption{The observed DAMA/LIBRA modulation spectrum from \cite{Bernabei:2022ath} as a function of electron-equivalent energy. The black bars show the $1\sigma$ confidence limits.}
		\label{fig:DAMAmodulationspectrum}
	\end{center}
\end{figure}

DAMA/LIBRA reports the modulation amplitude of the annual modulation signal binned in recoil energy, expressed in units of ${\rm keV_{ee}}$, with the subscript standing for electron equivalent energy. This means that the measured light yield corresponds to an electron recoil event with the given recoil energy. Therefore, if the scintillation event is due to a nuclear recoil, then the nuclear recoil energy is obtained as
\begin{equation}
    E_{\rm nr} = \frac{E_{\rm ee}}{Q\!F_T},
\end{equation}
where $E_{\rm ee}$ is the electron equivalent energy and $Q\!F_T$ is the quenching factor for the target nucleus $T$. We use the nominal values $Q\!F_{\rm Na}=0.3$ and  $Q\!F_{\rm I}=0.09$ \cite{Bernabei:2019ajy} in the following analysis. In section \ref{sec:resolution} we characterize the sensitivity of our results to a possible energy dependence of the $Q\!F$. The modulation amplitude observed by DAMA/LIBRA, binned in electron equivalent recoil energy is shown in figure \ref{fig:DAMAmodulationspectrum}. 

To assess the compatibility of COSINUS and DAMA/LIBRA within a given model, a straightforward procedure is to find the likelihood of obtaining the observed data from both COSINUS and DAMA/LIBRA, given the model parameters. Here our goal is instead to perform a model-independent comparison, for which this usual forward analysis is not possible.
Because COSINUS will directly measure the nuclear recoil energy, we would like to infer the nuclear recoil spectrum from the DAMA/LIBRA modulation spectrum, to allow a direct and model-independent comparison between the two experiments. An approximate solution to this \emph{inverse problem} can be obtained via a process of unfolding \cite{Zech:2012ch}; Let the true nuclear recoil spectrum be represented by a histogram with $N_{\rm bin}$ bins, $\{x_j\}$, $j=1,\ldots,N_{\rm bin}$, and let $A_{ij}$ be the probability for an event in the nuclear recoil energy bin $j$ to be observed in the electron equivalent energy bin $i$. Then, the expected number of observed events in bin $y_i$ is given by:
\begin{equation}
    y_i = \sum_{j=1}^{N_{\rm bin}} A_{ij}x_j.
\label{eq:folding}
\end{equation}
The inverse problem is to find an estimate for the nuclear recoil energy histogram $\{x\}$, given the observed histogram $\{y\}$. For this purpose, there exist various unfolding algorithms, see e.g. \cite{Schmitt:2016orm}. We use the iterative Richardson-Lucy unfolding algorithm described in \cite{Zech:2012ch}\footnote{We have also implemented the TUnfold algorithm \cite{Schmitt:2012kp}, and observed that it yields similar results for most nuclear recoil energy bins, but tends to return negative rates for the high energy bins. Using only the bins up to which the unfolded rate remains stable in TUnfold gives results within a few \% of those obtained with the iterative Richardson-Lucy algorithm.}. The algorithm produces an estimate for the event count in the nuclear recoil bin $x_j$ at the iteration step $k+1$ as
\begin{equation}
    x_j^{(k+1)}=\sum_{i=1}^M\frac{A_{ij}x_j^{(k)}y_i}{\sum_{l=1}^{N_{\rm bin}}A_{il}x_l^{(k)}\sum_{m=1}^M A_{mj}},
\end{equation}
where $x_j^{(k)}$ is the estimate at iteration step $k$. The prior distribution $\{x_j^{(0)}\}$ needs to be selected by hand. We have chosen a flat prior and verified that the results are not significantly affected by choosing a different smooth prior. A prior constructed by bin-to-bin correction factors (see \cite{Schmitt:2016orm}) yields results at most within 2\% of those obtained with the flat prior, and in most cases within less than 1\%. The algorithm will converge to a maximum likelihood solution, which often suffers from overfitting the noise in the data, resulting in strong anticorrelation between neighboring bins, as described in \cite{Zech:2012ch}. To regulate this behaviour, the number of bins in the true histogram $N_{\rm bin}$ should be smaller than the number of bins in the observed histogram $M$, and the algorithm should be stopped after a finite number of iteration steps, before it has converged to the maximum likelihood solution. The number of nuclear recoil energy bins $N_{\rm bin}$ and the number of iteration steps, $N_{\rm iter}$ are thus free parameters that need to be selected appropriately, but for which no simple uniquely defined criteria exist. We have varied these parameters in our analysis to confirm that the conclusions are not strongly dependent on a particular choice of the values of $N_{\rm bin}$ and $N_{\rm iter}$.

The response matrix $A$ for nuclear recoils off target nucleus $T$ is given by
\begin{equation}
A^T_{ij}=\frac{1}{E^{\rm max}_{{\rm nr}\, j}-E^{\rm min}_{{\rm nr}\, j}}\int_{E^{\rm min}_{{\rm nr}\, j}}^{E^{\rm max}_{{\rm nr}\, j}}\epsilon_{\rm DAMA}^T(E_{\rm nr};E^{\rm min}_{{\rm ee}\, i},E^{\rm max}_{{\rm ee}\, i}) dE_{\rm nr},
\end{equation}
where the efficiency function is \cite{Savage_2009}
\begin{equation}
\epsilon_{\rm DAMA}^T(E_{\rm nr};E_{\rm ee}^{\rm min},E_{\rm ee}^{\rm max}) = \frac12\left( {\rm erf}\left( \frac{E_{\rm ee}^{\rm max}-Q\!F_T E_{\rm nr}}{\sqrt{2}\sigma_{\rm DAMA}(Q\!F_TE_{\rm nr})} \right)-{\rm erf}\left( \frac{E_{\rm ee}^{\rm min}-Q\!F_T E_{\rm nr}}{\sqrt{2}\sigma_{\rm DAMA}(Q\!F_T E_{\rm nr})} \right) \right).
\label{eq:DAMAefficiency}
\end{equation}
This function describes the probability to observe a nuclear recoil event with recoil energy $E_{\rm nr}$ in the electron equivalent energy bin $[E_{\rm ee}^{\rm min},E_{\rm ee}^{\rm max}]$, assuming 
Gaussian spread.
The DAMA/LIBRA energy resolution function is given by \cite{Bernabei_2008}
\begin{equation}
\sigma_{\rm DAMA}(Q\!F_TE_{\rm nr}) = (0.448\ {\rm {keV}_{ee}})\sqrt{Q\!F_TE_{\rm nr}/{\rm {keV}_{ee}}}+0.0091Q\!F_TE_{\rm nr}.
\end{equation}
For the COSINUS resolution, we rely on the estimated baseline resolution $\sigma=0.2$ keV, and note that as long as the resolution is well below the bin width of the unfolded histogram, our results are not sensitive to the assumed COSINUS resolution.

\begin{figure}
	\begin{center}
		\includegraphics[width=0.49\linewidth]{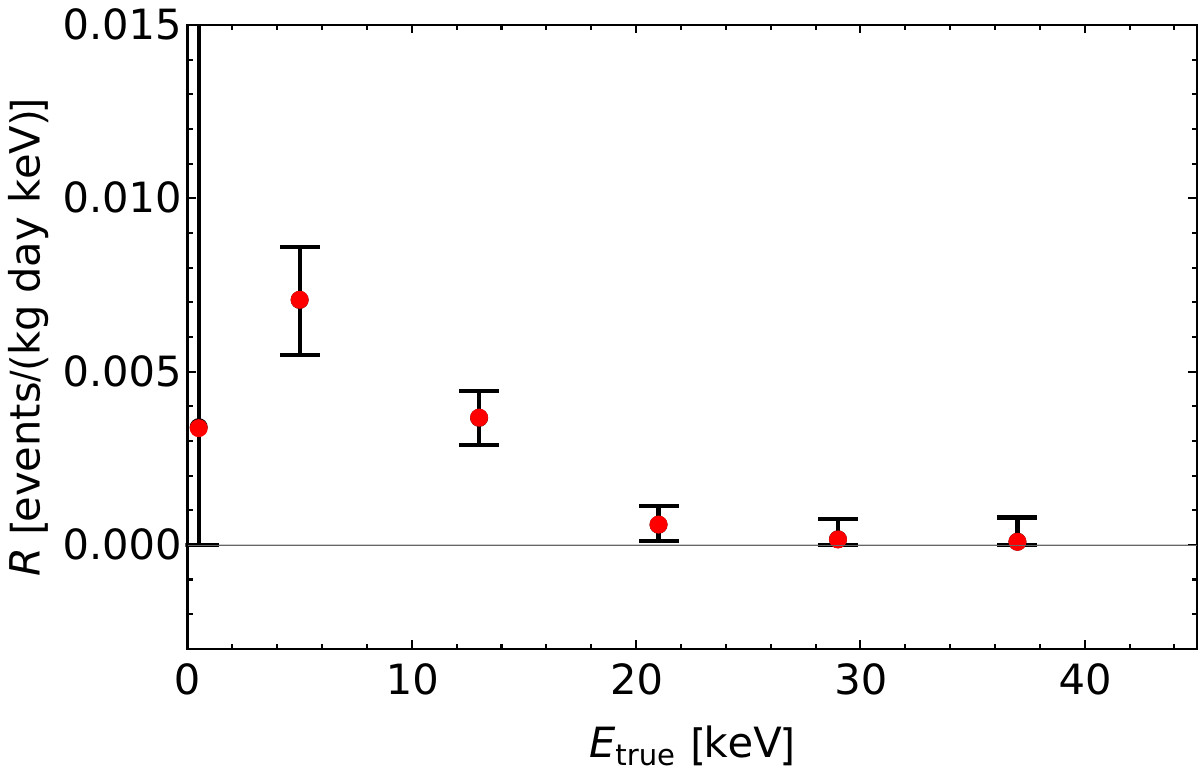}
        \includegraphics[width=0.49\linewidth]{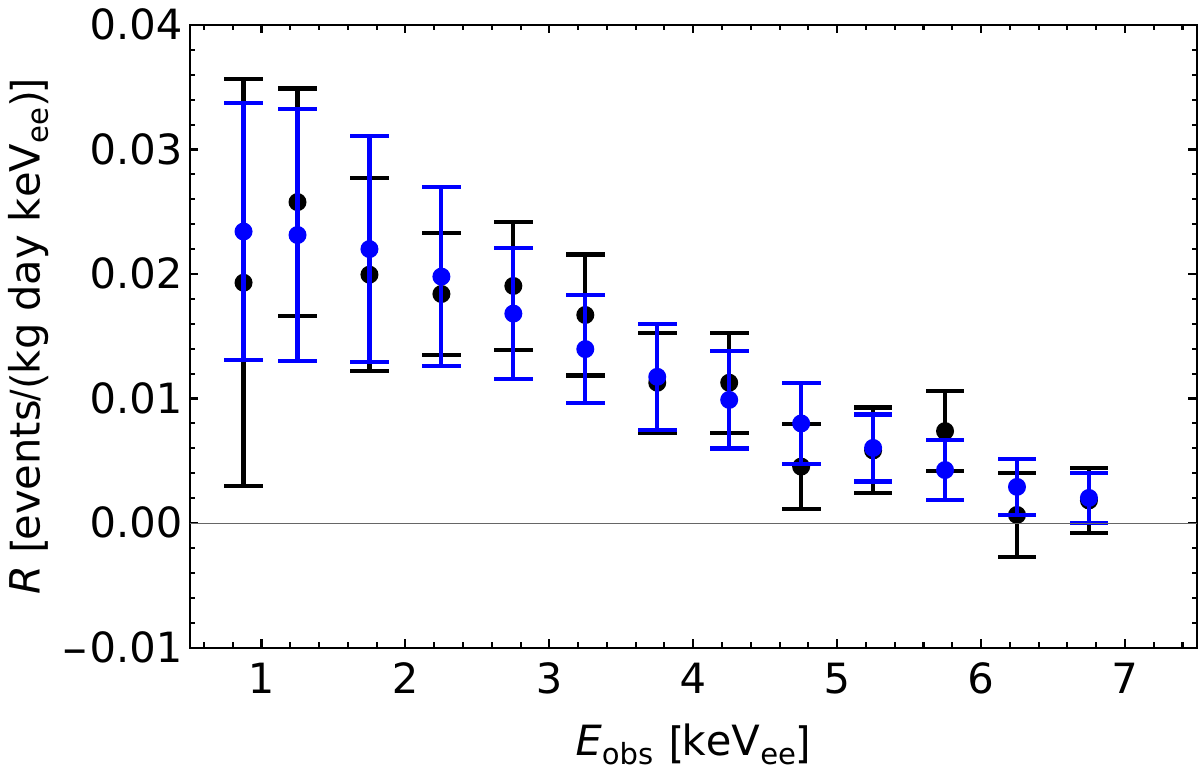}
		\caption{Left: the estimated nuclear recoil rate (red dots) obtained by unfolding the DAMA/LIBRA modulation amplitude data with $N_{\rm iter}=5$, assuming 1 keV energy threshold. The error bars show the $2\sigma$ confidence limits. The median values of the unfolded mock data samples used for the error estimate coincide with the unfolded nominal values shown by the red dots. Right: the expected DAMA/LIBRA modulation amplitude (blue dots) given the unfolded event rate shown in the left panel, overlaid with the actual DAMA/LIBRA data (black dots). The error bars are $2\sigma$ limits.}
		\label{fig:truespectrum}
	\end{center}
\end{figure}

We obtain an estimate for the binned nuclear recoil rate from the DAMA/LIBRA modulation data via the unfolding procedure described above. We have chosen the nuclear recoil energy bins so that the first bin is from zero to the assumed COSINUS detection threshold, for which we have used values ranging from 0.5 to 4 keV, and the rest of the bins have constant width and extend to $\sim 40$ keV for ${\rm Na}$ recoils and to $\sim 130$ keV for ${\rm I}$ recoils. To obtain an error estimate on the nuclear recoil spectrum we have generated a sample of 100k mock data sets for the DAMA/LIBRA modulation data, by randomly varying each modulation amplitude data point according to a Gaussian distribution with the standard
deviation given by the uncertainty reported in ref~\cite{Bernabei:2019ajy}, and shown in figure \ref{fig:DAMAmodulationspectrum}. From this sample of mock data sets, we obtain a sample of nuclear recoil spectra by the unfolding procedure. We then find the $1\sigma$, $2\sigma$ and $3\sigma$ confidence level upper and lower limits for the event rates in each nuclear recoil energy bin by considering the distribution of unfolded event rates for that bin in the generated sample. We find the lower limit at x\% confidence level for the event rate in the bin by selecting the value of the unfolded event rate which is smaller than $x$\% of the generated sample. An example of the resulting estimate for the binned nuclear recoil event rate with $2\sigma$ confidence level regions is shown in the left panel of figure \ref{fig:truespectrum}. The estimated event rate, obtained by unfolding the DAMA/LIBRA data is shown by the red dots. The black error bars show the $2\sigma$ confidence region of the unfolded mock data sets. The right panel of the figure shows the expected DAMA/LIBRA modulation amplitudes if the nuclear recoil event rate was given by the estimated nuclear recoil histogram (see eq (\ref{eq:folding})), overlaid with the actual DAMA/LIBRA data. 

\begin{figure}
	\begin{center}
		\includegraphics[width=0.5\linewidth]{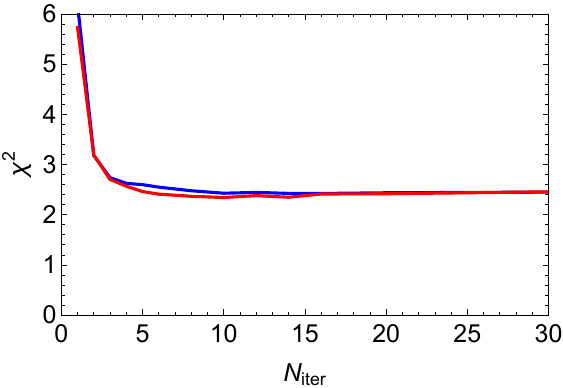}
		\caption{The $\chi^2$ value for the forward folded spectrum compared to the observed DAMA/LIBRA modulation spectrum as a function of the number of iteration steps in the unfolding procedure. The blue (red) line is for 3 (5) bins in the estimated true energy spectrum.}
		\label{fig:convergence}
	\end{center}
\end{figure}

As discussed above, the iterative Richardson-Lucy algorithm converges to the maximum likelihood estimate for the true spectrum, which may feature large anticorrelations between neighboring bins. This unphysical feature can be avoided by reducing the number of bins $N_{\rm bin}$ in the true spectrum and by reducing the number of iteration steps $N_{\rm iter}$. To find an appropriate number of iteration steps, we have tested the convergence of the forward model, i.e. the expected spectrum obtained by folding back the unfolded spectrum as shown in the right panel of figure \ref{fig:truespectrum}. In figure \ref{fig:convergence}, we show the $\chi^2$-value\footnote{We use the $\chi^2$-test as defined for independent variables to test the convergence of the forward folded spectrum. While the maximal likelihood estimate for the unfolded spectrum may contain strong (anti)correlations, this is not the case for the forward folded spectrum. We have checked that using a quadratic form for the $\chi^2$ parameter with a covariance matrix obtained from the sampled mock data sets results in nearly identical curve.} for the difference of the observed DAMA/LIBRA modulation amplitude $R^{\rm DAMA}$ with the forward folded mean values $R^{\rm mean}$ obtained after $N_{\rm iter}$ iterations,
\begin{equation}
    \chi^2 = \sum_i\frac{(R_i^{\rm DAMA}-R_i^{\rm mean})^2}{\sigma_{R_i}^2},
\end{equation}
where the sum is over the DAMA/LIBRA modulation bins shown in figure \ref{fig:DAMAmodulationspectrum}, and the error $\sigma_{R_i}$ is obtained by combining the errors of the DAMA/LIBRA modulation amplitudes and the forward folded amplitudes in quadrature. We observe that the compatibility of the forward folded and the actual observed data sets quickly converges after a few iteration steps, and does not significantly improve after further iterations. Therefore we choose $N_{\rm iter}=5$ iteration steps for the analysis. Unless mentioned otherwise, this number will be used hereafter.

To perform the comparison between the anticipated COSINUS data and the unfolded DAMA/LIBRA modulation data, we make use of the notion that the modulation amplitude cannot exceed the total DM event rate \cite{Kahlhoefer:2018knc}. Therefore, the unfolded rate, which was obtained from the modulation amplitude histogram, represents a lower limit on the DM event rate that could explain the DAMA/LIBRA annual modulation signal. For now, we assume that COSINUS operates at zero background level and observes no events in the region of interest, resulting in upper limit on the event rate in the region of interest, using Poisson statistics e.g. for a one-sided $2\sigma$-limit, 
\begin{equation}
    R_{\rm bound}=\frac{3.78}{\epsilon_{\rm COSINUS}\mathcal{E}},
\label{eq:Rbound}
\end{equation}
where $\epsilon_{\rm COSINUS}$ is the detection efficiency of the COSINUS experiment (assumed to be constant over the region of interest) and $\mathcal{E}$ is the exposure. We use the estimate $\epsilon_{\rm COSINUS}=0.38$ for sodium recoils and $\epsilon_{\rm COSINUS}=0.76$ for iodine recoils, where the difference is due to event selection below the median light yield of the sodium recoil band. The base acceptance of the detector is assumed as 76\%. For further details see \cite{Angloher:2025shf}. 

From the unfolded histogram we combine the bins above the assumed COSINUS detection threshold that indicate a non-zero event rate, combining the errors in quadrature. If the resulting $n\sigma$ lower limit on the event rate is larger than the COSINUS $n\sigma$ upper limit $R_{\rm bound}$, we conclude that the DM explanation for the DAMA/LIBRA annual modulation signal is excluded at the $n\sigma$ confidence level. We repeat this analysis for varying values of the COSINUS detection threshold, and for varying choices for the number of nuclear recoil energy bins $N_{\rm bin}$ in the unfolding algorithm. For each selection of parameters, we determine the required COSINUS exposure by equating $R_{\rm bound}$ to the lower limit of the unfolded event rate. The results are shown in table \ref{tab:exposure}.

\begin{table}[h!]
    \centering
    \begin{tabular}{c|c|c|c|c|c|c|c}
          &  &  ${\rm Na}$ & ${\rm Na}$  & ${\rm Na}$ &  ${\rm I}$ &  ${\rm I}$ &  ${\rm I}$ \\
        $E_{\rm min}$ & $N_{\rm bin}$ &$\mathcal{E}(3\sigma)$ &$ \mathcal{E}(2\sigma)$ & $\mathcal{E}(1\sigma)$  & $\mathcal{E}(3\sigma)$  & $\mathcal{E}(2\sigma)$ & $\mathcal{E}(1\sigma)$\\
        keV &  &   kg day  & kg day & kg day & kg day & kg day & kg day \\ \hline
       0.5 & 3 & 230.6 & 120.9 & 54.4 & 112.1 & 59.0 & 26.5 \\
       0.5 & 5 & 240.3 & 123.8 & 54.8 & 119.3 & 60.9 & 26.8 \\
       0.5 & 7 & 248.3 & 125.4 & 55.2 & 121.8 & 61.7 & 27.0 \\
       0.5 & 9 & 249.6 & 127.7 & 55.7 & 123.8 & 62.9 & 27.3 \\ \hline
       1 & 3 &  240.8 & 125.8 & 56.5 & 113.2 & 59.7 & 26.9 \\
       1 & 5 &  244.2 & 127.6 & 56.8 & 119.2 & 61.7 & 27.2 \\
       1 & 7 &  252.0 & 129.5 & 57.1 & 122.5 & 61.7 & 27.2 \\
       1 & 9 &  255.5 & 130.1 & 57.3 & 124.5 & 63.7 & 27.6 \\ \hline
       2 & 3 &  256.6 & 135.9 & 61.3 & 116.3 & 61.1 & 27.5 \\
       2 & 5 &  261.5 & 135.8 & 60.9 & 122.9 & 62.8 & 27.7 \\
       2 & 7 &  260.1 & 136.0 & 60.8 & 125.1 & 63.0 & 27.8 \\
       2 & 9 &  262.8 & 136.3 & 60.9 & 127.1 & 64.3 & 28.0 \\ \hline
       3 & 3 &  281.2 & 150.7 & 68.1 & 118.6 & 62.4 & 28.0 \\
       3 & 5 &  278.1 & 146.8 & 66.4 & 123.0 & 63.5 & 28.3 \\
       3 & 7 &  280.4 & 146.4 & 65.9 & 125.4 & 64.2 & 28.3 \\
       3 & 9 &  279.8 & 146.3 & 65.8 & 127.5 & 64.9 & 28.5 \\ \hline
       4 & 3 &  317.4 & 169.4 & 77.0 & 122.3 & 64.0 & 28.7 \\
       4 & 5 &  310.0 & 162.9 & 73.6 & 126.0 & 64.7 & 28.9  \\
       4 & 7 &  311.1 & 161.5 & 72.7 & 126.9 & 65.2 & 28.9 \\
       4 & 9 &  304.5 & 160.9 & 72.6 & 128.3 & 65.9 & 29.0 \\
    \end{tabular}
    \caption{The required exposure, assuming zero background in the event selection region, for $3\sigma$, $2\sigma$ and $1\sigma$ confidence exclusion of the DAMA/LIBRA nuclear recoil DM signal using various binning of the estimated spectrum in the unfolding procedure, for both elements. The parameter $N_{\rm bin}$ refers to the number of bins in the visible region in the unfolded histogram, so that the total number of bins in the nuclear recoil histogram is $N_{\rm bin}+1$. The detection threshold assumed for COSINUS is $E_{\rm min}$.}
    \label{tab:exposure}
\end{table}

We observe that the results are robust with respect to variations in the $N_{\rm bin}$ parameter. The required exposure grows slowly with the assumed detection energy threshold, as expected. We will use the coarser binning, with $N_{\rm bin} = 3 $ or $N_{\rm bin} = 5$ hereafter, unless otherwise specified. It is also evident that if the scattering is off iodine nuclei, the comparison is always more favorable for COSINUS than in the case of scattering off sodium. This is due to both the smaller quenching factor of iodine, which means that a smaller part of the signal rate can be below the COSINUS detection threshold, and due to the higher detection efficiency of iodine recoil events in COSINUS, due to the smaller light yield. Therefore, in what follows we will mostly present results assuming 100\% scattering off sodium, as this will result in conservative estimates for the required exposure. If any part of the DAMA/LIBRA signal comes from iodine recoils, this part will be more easily observable in COSINUS and therefore the exclusion can be reached with less exposure in that case.

To estimate the effect of the COSINUS background event rate on the required exposure, we perform an analysis with a simple background model. We assume a flat electron recoil background level of 1 event per keV kg day, and a $^{40}\rm{K}$ contamination corresponding to an activity of 600 $\mu{\rm Bq}/{\rm kg}$, as described in \cite{Angloher:2016ooq}. We then use the band model presented in \cite{COSINUS:2023kqd} to define the event selection region below the median of the $\rm{Na}$ recoil band, and model the leakage of the electron recoil events to the selection region assuming light detector energy resolution of $\sigma_{\rm L}=0.11\ {\rm keV_{ee}}$. For further details of the background model see \cite{Angloher:2025shf}. We then find the expected upper limit for the event rate in the presence of this background, by computing the expectation value of the upper limit $\langle R_{\rm bound} \rangle$ using Poisson statistics for the observed number of background events, following the procedure presented in \cite{Roe:2000fe}. This expected upper limit is then compared to the lower limit of the event rate obtained from the unfolding procedure, as explained above.

\begin{figure}
	\begin{center}
		\includegraphics[width=0.8\linewidth]{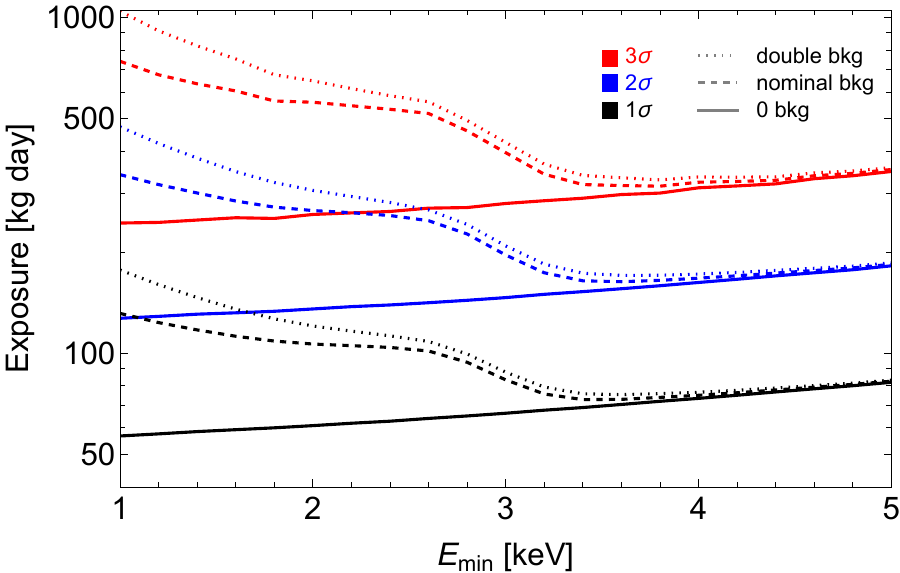}
		\caption{The required exposure for $1\sigma$ (black), $2\sigma$ (blue) and $3\sigma$ (red) exclusion of the DAMA/LIBRA dark matter signal as a function of $E_{\rm min}$ for $N_{\rm bin}=5$, $N_{\rm iter}=5$. The solid lines are for zero background, the dashed lines for the nominal background model, and the dotted lines for a background model with doubled rate in the flat component.}
		\label{fig:exposure_threshold}
	\end{center}
\end{figure}

We show the resulting required exposure for a $1\sigma$, $2\sigma$ and $3\sigma$ exclusion of the DAMA/LIBRA nuclear recoil signal as a function of the COSINUS event selection threshold in figure \ref{fig:exposure_threshold}, for the case of zero background (solid lines), the nominal background model described above (dashed lines), and a background model where the flat background component has been doubled but the $^{40}\rm{K}$ contamination kept at its nominal value. We observe that for low threshold the leakage may have a significant effect, but this can be avoided by restricting the event selection above 3 keV. We postpone a detailed analysis of the optimal event selection strategy in the presence of a more complete background model for a future publication.

\section{Effect of energy resolution and quenching factor}
\label{sec:resolution}

In the above analysis, we have used the nominal values for the DAMA/LIBRA energy resolution and quenching factors, as reported by the DAMA/LIBRA collaboration.
We will now investigate how our results depend on the assumed values of these parameters in the unfolding procedure, in case they would differ from the reported values.

\begin{figure}
	\begin{center}
		\includegraphics[width=0.49\linewidth]{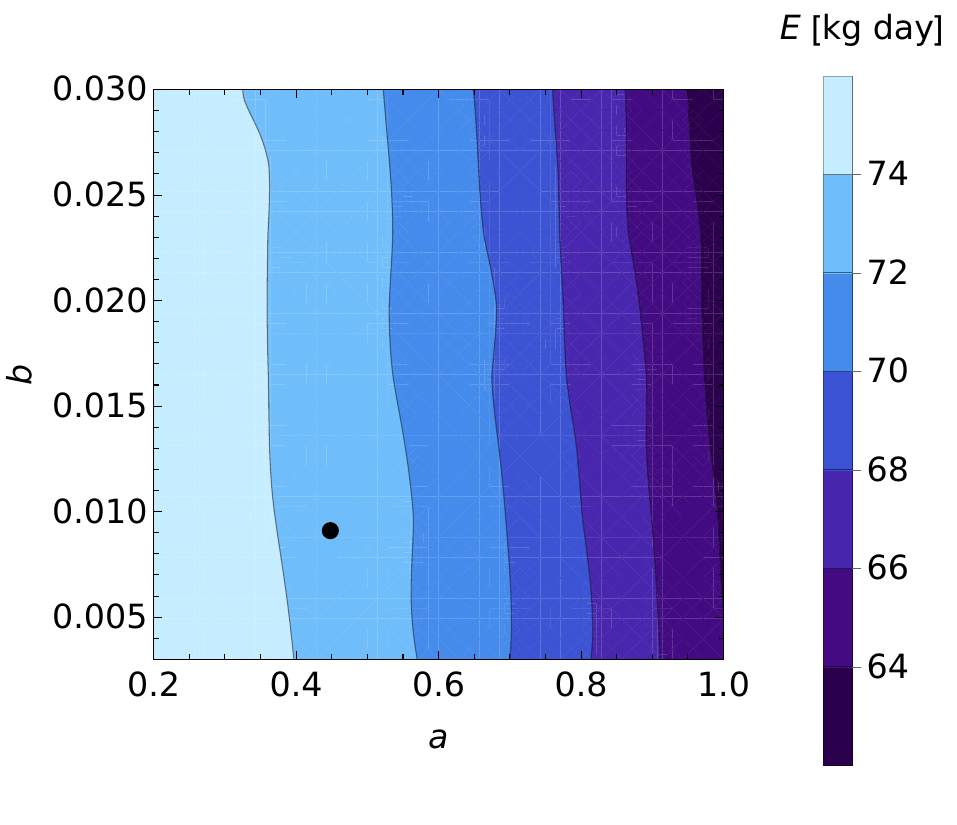}
        \includegraphics[width=0.49\linewidth]{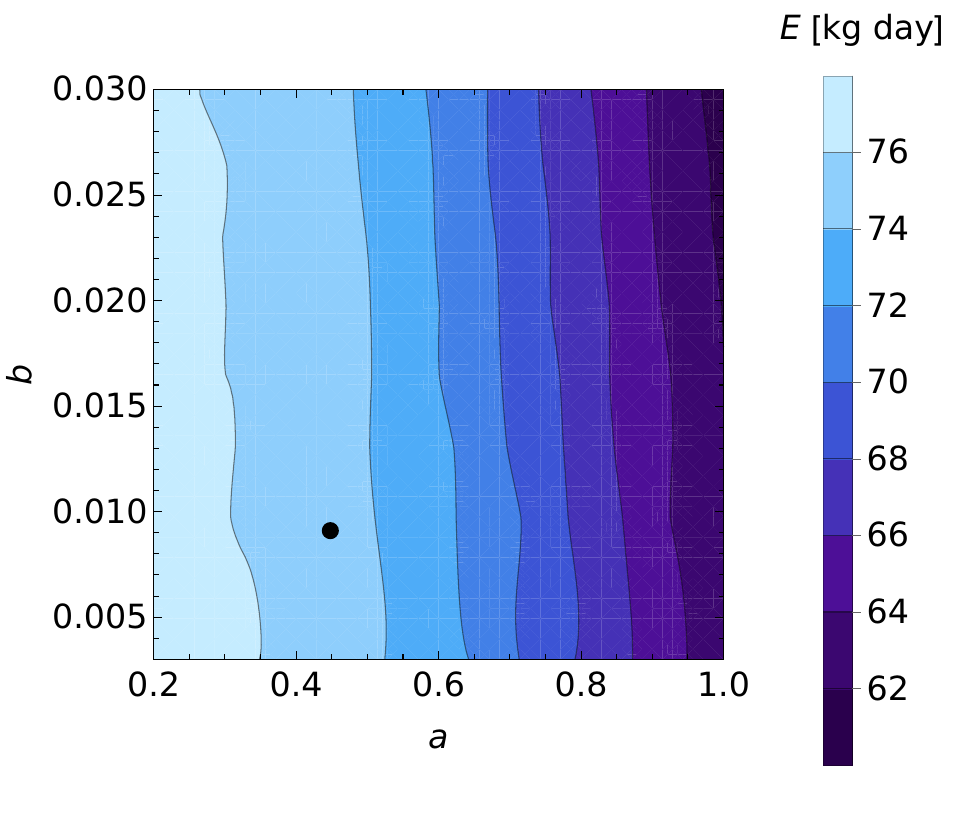}
		\caption{The required exposure for 90\% CL rejection of the DAMA/LIBRA dark matter signal, for $E_{\rm min}=1$ keV, $N_{\rm bin}=3,5$ (left, right) and $N_{\rm iter}=5$, as a function of the resolution function parameters $a,b$ in the parametrization (\ref{eq:resolution}). The nominal values for the parameters are shown by the black dot.}
		\label{fig:Resolutionfunction}
	\end{center}
\end{figure}

We parametrize the DAMA/LIBRA resolution function as
\begin{equation}
\sigma_{\rm DAMA}(Q\!F_TE_{\rm nr}) = (a\ {\rm {keV}_{ee}})\sqrt{Q\!F_TE_{\rm nr}/{\rm {keV}_{ee}}}+bQ\!F_TE_{\rm nr},
\label{eq:resolution}
\end{equation}
where the nominal values for the parameters are $a=0.448$, $b=0.0091$ \cite{Bernabei_2008}. In figure \ref{fig:Resolutionfunction}, we show how the required exposure for COSINUS depends on these parameters, using 3 (left) or 5 (right) bins in the unfolding procedure, with $N_{\rm iter}=5$ and $E_{\rm min}=1$ keV. These results were obtained by repeating the analysis procedure described in section \ref{Analysis} for varying values of the resolution function parameters $a$ and $b$. We observe that the required exposure is rather insensitive to the assumed DAMA/LIBRA resolution,  varying by $\sim \pm 10\%$ within the scanned parameter range, so that the conclusions of the above analysis hold even if there is some uncertainty on the true energy resolution.

\begin{figure}
	\begin{center}
		\includegraphics[width=0.49\linewidth]{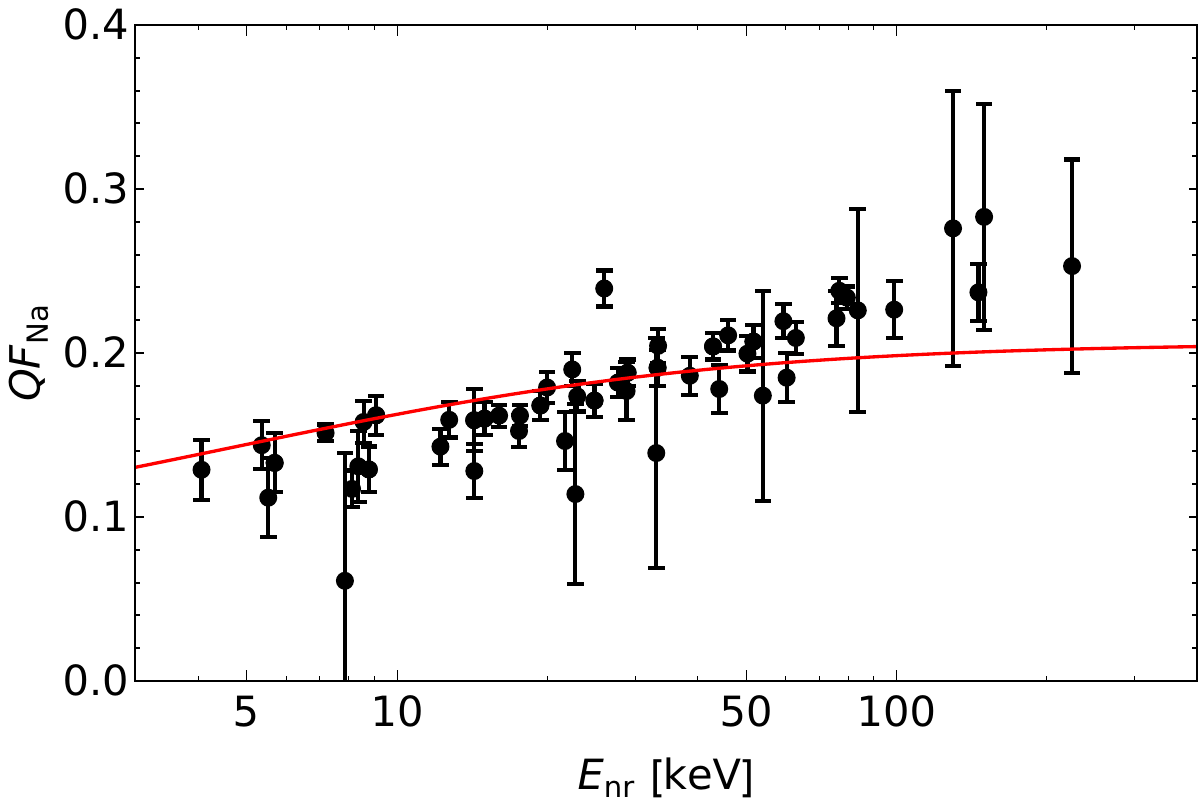}
        
		\caption{A fit with $\chi^2/{\rm dof.}=5.6$ to the quenching factor data \cite{Xu:2015wha,Collar:2013gu,Stiegler:2017kjw,Bharadwaj:2023aoz,Cintas:2024pdu,Lee:2024unz} with the parametric form (\ref{eq:QFfit}), with $\alpha=0.213$, $\beta=0.206$.}
		\label{fig:QFfit}
	\end{center}
\end{figure}

To account for a possibly energy-dependent quenching factor for ${\rm Na}$ in DAMA/LIBRA, we use a parametrization motivated by the Lindhard model \cite{Chagani:2008in}:
\begin{equation}
    Q\!F_{\rm Na}(E_{\rm nr})= \beta \frac{\alpha g(E_{\rm nr})}{1 + \alpha g(E_{\rm nr})}, \quad g=3E_{\rm nr}^{0.15} + 0.7E_{\rm nr}^{0.6} + E_{\rm nr}.
    \label{eq:QFfit}
\end{equation}
Here, the parameters $\alpha$ and $\beta$ control the shape and normalization of the $Q\!F$, respectively. A fit of this form to a sample of measured values for the quenching factor \cite{Xu:2015wha,Collar:2013gu,Stiegler:2017kjw,Bharadwaj:2023aoz,Cintas:2024pdu,Lee:2024unz} is shown in figure \ref{fig:QFfit}. We observe that this parametrization is not a very good fit to the data, with the best-fit $\chi^2/{\rm dof.}=5.6$. However, the purpose of this exercise is simply to characterize the dependence of our results on the possibility that the DAMA/LIBRA $Q\!F$s might differ from the constant values used in the analysis, and could depend on the recoil energy in some way. Therefore the particular form of the energy dependence is not our main interest here. We only consider the energy dependence of the ${\rm Na}$ $Q\!F$ here, since the scenario in which most of the scattering is off Iodine is, in any case, more favorable for COSINUS, as is evident from table~\ref{tab:exposure}. 

\begin{figure}
	\begin{center}
		\includegraphics[width=0.49\linewidth]{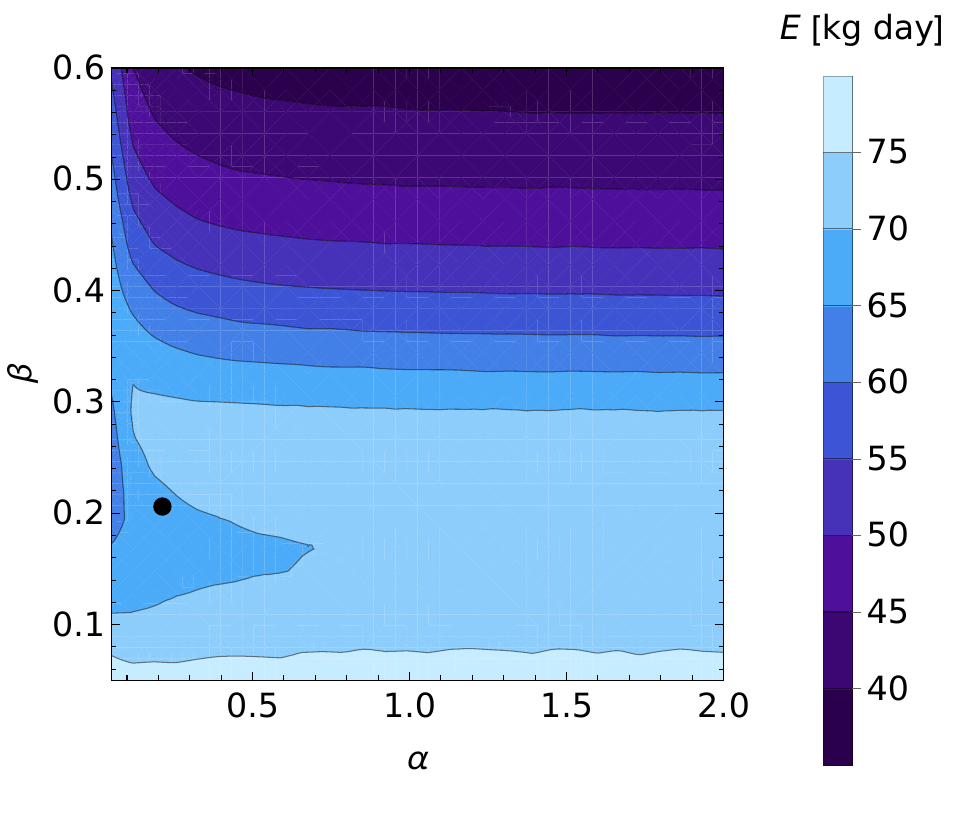}
        \includegraphics[width=0.49\linewidth]{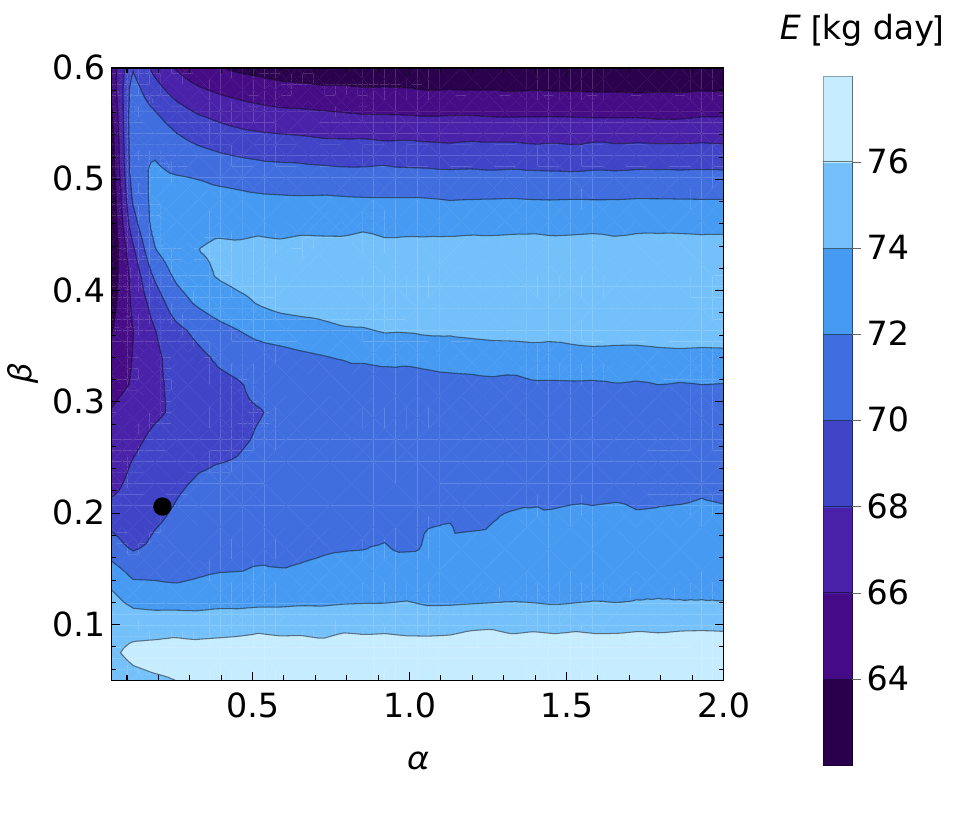}
        
		\caption{The required exposure for 90\% CL rejection of the DAMA/LIBRA dark matter signal, for $E_{\rm min}=1$ keV, $N_{\rm bin}=3,5$ (left, right) and $N_{\rm iter}=5$, as a function of the quenching factor parameters $\alpha,\beta$ using the parametrization (\ref{eq:QFfit}). The best fit values for the parameters are shown by the black dot. The DAMA/LIBRA nominal values correspond to $\alpha=\infty$, $\beta=0.3$.}
		\label{fig:QFscan}
	\end{center}
\end{figure}

We then vary the parameters $\alpha,\beta$ in eq. (\ref{eq:QFfit}) and repeat the analysis procedure described in section \ref{Analysis} with $N_{\rm iter}=5$, $E_{\rm min}=1$ keV. The resulting required exposure for a 90\% CL exclusion of the DAMA/LIBRA DM signal is shown in figure~\ref{fig:QFscan}. We again notice that the result is not very sensitive to changes in the quenching factor for realistic values of the Lindhard model parameters. Therefore our results can be deemed reliable regardless of the uncertainty about the DAMA/LIBRA quenching factors.

 This insensitivity to the variation in the $Q\!F$ is largely due to the assumed 1 keV energy threshold and zero background above the threshold. Therefore, we also comment on a hypothetical worst case scenario, where COSINUS would be background dominated in the low energy region, so that the effective event selection threshold would be as high as 4 keV. If, simultaneously, the DAMA/LIBRA quenching factor were abnormally high, this could lead to a large portion of the DAMA/LIBRA signal being hidden below the COSINUS threshold. In an extreme scenario with $Q\!F_{\rm Na}=1$, $E_{\rm min}=4\ {\rm keV}$, we obtain the required exposure to be 240 kg day, 700 kg day and 2300 kg day for $1\sigma$, $2\sigma$ and $3\sigma$ confidence level exclusions, respectively. Although unrealistic, this worse case scenario serves to highlight the importance of reaching a sufficiently low energy threshold in COSINUS, to be able to see the DAMA/LIBRA signal even in the case of an abnormally high quenching factor hypothesis.

\section{Conclusions}

In the context of a given dark matter model, the results of any two experiments can be compared by fitting the model to both data sets, and evaluating the goodness of fit to the combination of data. However, for a model-independent analysis, such forward modeling is not possible. Therefore, we have performed an unfolding of the dark matter event rate implied by the DAMA/LIBRA annual modulation data in order to estimate the sensitivity of the COSINUS experiment to such event rate. We find that, in an optimistic zero-background scenario and a 1 keV nuclear recoil detection threshold, COSINUS can reach a $1\sigma$ exclusion of the DAMA/LIBRA DM signal with 57 kg day exposure. For $2\sigma$ exclusion the required exposure is 130 kg day, and for $3\sigma$ exclusion, 250 kg day. Assuming gaussian errors for the unfolded distributions, we estimate the required exposure for a $5\sigma$ exclusion to be 750 kg day. A simple background model indicates that the low energy region below $\sim3.5$ keV could be contaminated by electron recoil events leaking into the nuclear recoil selection window. If the event selection is restricted above 4 keV, we find the required exposure as 77 kg day ($1\sigma$), 170 kg day ($2\sigma$) and 320 kg day ($3\sigma$).
We have investigated the sensitivity of these results to assumptions made about the DAMA/LIBRA quenching factor and energy resolution, and found them to be stable under reasonable variations in these parameters.

\section*{Acknowledgements}We thank Felix Kahlhoefer for discussions at early stages of this work. Financial support from the Research Council of Finland (grant\# 342777), Vilho, Yrjö and Kalle Väisälä fund, Austrian Science Fund FWF (grant DOI 10.55776/PAT1239524) and Klaus Tschira Foundation is gratefully acknowledged.

\bibliography{bibliography.bib}

\end{document}